\documentclass[12pt]{article}
\usepackage{amsfonts,amssymb,amsmath}

\vspace{20mm}

\let\ddpp\mathbb 
\def\R{\ddpp R}     

\newtheorem{thm}{Theorem}[section]
\newtheorem{cor}[thm]{Corollary}

\newtheorem{prop}[thm]{Proposition}

\newtheorem{rem}[thm]{Remark}
\newtheorem{ex}[thm]{Example}

\begin{document}

\title{The Gauss-Landau-Hall problem on Riemannian
surfaces}

\author{Manuel Barros$^1$, Jos\'{e} L. Cabrerizo$^2$, Manuel Fern\'{a}ndez$^2$ \\ and Alfonso
Romero$^1$ \\[3mm] ${}^1$Departamento de Geometr\'ia y Topolog\'ia, Facultad de Ciencias \\Universidad
de Granada, 18071-Granada, Spain. \\E-mail addresses: {\ttfamily
mbarros@ugr.es, aromero@ugr.es}
\\[3mm] ${}^2$Departamento de Geometr\'ia y Topolog\'ia, Facultad
de Matematicas \\ Universidad de Sevilla, 41012-Sevilla, Spain.
\\E-mail addresses: {\ttfamily jaraiz@us.es, mafernan@us.es}}

\date{}
\maketitle

\begin{abstract}
We introduce the notion of Gauss-Landau-Hall magnetic field on a
Riemannian surface. The corresponding Landau-Hall problem is shown
to be equivalent to the dynamics of a massive boson. This allows
one to view that problem as a globally stated, variational one. In
this framework, flowlines appear as critical points of an action
with density depending on the proper acceleration. Moreover, we
can study global stability of flowlines. In this equivalence, the
massless particle model correspond with a limit case obtained when
the force of the Gauss-Landau-Hall increases arbitrarily. We also
obtain new properties related with the completeness of flowlines
for a general magnetic fields. The paper also contains new results
relative to the Landau-Hall problem associated with a uniform
magnetic field. For example, we characterize those revolution
surfaces whose parallels are all normal flowlines of a uniform
magnetic field.
\end{abstract}


\section{From a classical picture to a general setting}

Classically, the Landau-Hall problem consists of the motion study
of a charged particle in the presence of a static magnetic field,
$H$. In this setting, free of any electric field, a particle, of
charge $e$ and mass $m$, evolves with velocity $v$ satisfying the
Lorentz force law, \cite{Landau-Lifshitz},

\[\frac{dP}{dt}=\frac{e}{c}\; v \times H,\]

\vspace{4mm}

\noindent where $c$ denotes the light speed, $P=(\epsilon/c^2)\,
v$ stands for the momentum of the particle, and $\epsilon =
mc^2\left[ 1- (\parallel v
\parallel^2/c^2)\right] ^{-1/2}$ is its energy. Since $dP/dt$ is
orthogonal to $P$, then $(d/dt) (\| P \|^2)=0$. This implies the
constancy of both $\| v\|$ and $\epsilon$. Assume $H$ is
stationary, i.e., $H$ is a time-independent vector of the
Euclidean space $\R ^3$. With the choice of a suitable orthonormal
reference system, we may assume that $H=h \,(0,0,1)$, for some $h
\in\mathbb{R}$. In this framework, we have
$$\frac{d}{dt}\,v_1(t)=\omega \, v_2(t), \quad
\frac{d}{dt}\,v_2(t)=-\omega \, v_1(t),\quad
\frac{d}{dt}\,v_3(t)=0,$$ \noindent where $\omega=(ehc)/\epsilon$
is constant. Then

\[x_1(t)=x_1^0 + r \sin(\omega t + \alpha), \quad x_2(t)=x_2^0 + r
\cos(\omega t + \alpha), \quad x_3(t)=x_3^0 + v_3^0t,\]

\noindent where $r=\| v\| /\omega$. In particular, if $v_3^0=0$,
then the particle describes a circle in the plane $x_3=x_3^0$,
with center $(x_1^0,x_2^0,x_3^0)$ and radius $r$. Now, in this
plane we consider the $2$-form $F$ defined by
$F(X,Y)=\varepsilon<X\times Y,H>$, where $\varepsilon=\pm 1$ is
the sign of $h/\omega$. It is clear that $F$ is covariantly
constant, and therefore it is a constant multiple of the area
element, indeed $F=\varepsilon h\, dx_1\wedge dx_2$. Now, consider
the metric $g$ on the plane defined by
$g:=\varepsilon(h/\omega)g^0$, where $g^0=< \, ,>$ denotes the
Riemannian metric on the plane induced by the usual one of
$\mathbb{R}^3$. Define the operator $\Phi$, $g$-equivalent to $F$,
by $g(\Phi(X),Y)=F(X,Y)$. Then, the Lorentz force law can be
expressed in terms of this form by

\begin{equation}\label{ecuacionenr3}
    \frac{d}{dt}\,v(t)=\, \Phi(v(t)).
\end{equation}

\vspace{4mm}

This approach to the classical picture can be obviously extended
to a more general setting. In fact, it seems natural to define a
{\it magnetic field} on a $n(\geq 2)$-dimensional Riemannian
manifold $(M,g)$, as a closed $2$-form $F$ on $M$. The {\it
Lorentz force} of a magnetic background $(M,g,F)$ is defined to be
the skew-symmetric operator, $\Phi$, given by

\begin{equation}\label{operator}
g(\Phi(X),Y)=F(X,Y),
\end{equation}

\vspace{2mm}

\noindent for any couple of vector fields $X,Y$ on $M$. Let us
remark that $\Phi$ is metrically equivalent to $F$, so no
information is lost when $\Phi$ is considered instead $F$. In
classical terminology, it is said that $\Phi$ is obtained from $F$
by raising its second index, and $\Phi$ and $F$ are then said to
be physically equivalent. On the other hand, there exists another
operator $\Phi'$ defined from $F$ via $g$ in a similar way, namely
$g(X,\Phi'(Y))=F(X,Y),$ but it is easily seen that $\Phi'=-\Phi$.
So, the choice from among $\Phi$ or $\Phi'$ to represent $F,$
using $g$, is not relevant. Along this paper, we will use $\Phi$
to denote the Lorentz force induced from $(M,g,F)$.

\vspace{2mm}

A (smooth) curve $\gamma$ in $(M,g)$ is called a {\it flowline} of
the dynamical system associated with the magnetic field $F$ (or
simply a flowline of $F,$ or a {\it magnetic curve} of $(M,g,F)$),
if its velocity vector field, $\gamma^{\prime}$, satisfies the
following (Landau-Hall) differential equation,

\[  \nabla_{\gamma^{\prime}} \gamma^{\prime}=\Phi(\gamma^{\prime}),
\eqno{\rm(LH)} \]

\vspace{2mm}

\noindent where $\nabla$ is the Levi-Civita connection of $g$
[compare with Eq. (\ref{ecuacionenr3})].

\vspace{4mm}

For the trivial magnetic field, $F=0$, the case without the force
of a magnetic field, magnetic curves correspond with the geodesics
of $(M,g)$. As it is well known, they are nicely characterized as
critical points of an energy action and so they represent the
trajectories for free fall particles (moving under the influence
of only gravity). In the general case, however, magnetic flows are
important examples of dynamical systems on Riemannian manifolds
whose flowlines, being the trajectories of charged particles in
(non trivial) magnetic fields, are not geodesics (Proposition
\ref{homogeneizacion}) but, as we will see later, they are closely
related with the Riemannian structure.

\vspace{2mm}

Nevertheless, the magnetic curves of $(M,g,F)$ can be also viewed,
at least locally, as the solutions of a variational principle. In
fact, let $U$ be an open subset of $M$ where $F=d\omega$ for some
potential $1$-form $\omega$ (this open subset could be the whole
$M$ when $H^{2}(M)=0$). For any two fixed points $p,q\in U$, we
consider the space $\Gamma_{pq}$ of smooth curves in $U$ that
connect these two points. Now, we choose the action
$\mathcal{LH}:\Gamma_{pq}\rightarrow\mathbb{R}$ defined by

\begin{equation} \label{primerfuncional}
\mathcal{LH}(\gamma)=\displaystyle \frac{1}{2}\int_{\gamma}
g(\gamma',\gamma')dt-\int_{\gamma} \omega(\gamma ')dt.
\end{equation}

\noindent The tangent space of $\Gamma_{pq}$ in $\gamma$ is made
up of the smooth vector fields, $V$, along $\gamma$ that vanish at
the end points $p,q\in U$. An standard computation involving
integration by parts allows one to compute the first variation of
this action to be

\[\delta(\mathcal{LH})(\gamma)[V]=\displaystyle -\int_{\gamma} g\left(
\nabla_{\gamma^{\prime}} \gamma^{\prime}-\Phi(\gamma
'),V\right)dt.\]

\noindent As a consequence, we get
\[\delta(\mathcal{LH})(\gamma)[V]=0, \; \mathrm{for \; any}
\; V\in T_{\gamma}\Gamma_{pq}  \; \; \mathrm{if \; and \; only \;
if \; \gamma \; is \;a \; solution \; of \; (LH)}.\]

\noindent This argument shows that the differential equation (LH)
is indeed the Euler-Lagrange equation associated with the
functional $\mathcal{LH}$.

\vspace{2mm}

However, it seems natural to realize the old idea of
characterizing magnetic curves from a global variational
principle. In other words, to obtain the magnetic trajectories of
$(M,g,F)$ as solutions of a variational problem that neither it
does not involves any local potential nor it does not constraint
the topology of $M$. This is, in general, an interesting open
problem. One of the main aim of this paper is just to solve it for
certain magnetic fields on surfaces.

\vspace{2mm}

To be precise, we introduce the notion of a Gauss-Landau-Hall
magnetic field (in brief, GMF) on an oriented Riemannian surface
$(M,g)$. First, we do it in the natural context that surfaces are
immersed in Euclidean space $\mathbb{R}^3$ using the Gauss map.
However, we notice that the notion of GMF is absolutely intrinsic
so it can be considered on surfaces even if they are not regarded
in $\mathbb{R}^{3}$. Then, we are able to obtain an amazing result
which characterizes the normal flowlines of a GMF as the solutions
of a variational principle globally stated. Therefore, those
flowlines appear as critical points of an action whose Lagrangian
density involves the proper acceleration of particles
(relativistic particles with rigidity of order one, in the sense
of Plyushchay, \cite{Plyushchay2},\cite{Plyushchay}). A priori,
these actions describe a massive relativistic boson. However,
massless particles with arbitrary helicity are obtained as a limit
case, just when the Lorentz force of the GMG increases
arbitrarily.

\vspace{2mm}

Other details on the paper are the following. We first provide in
Section 2 an analysis of the existence, uniqueness, extendibility
and completeness of the magnetic curves associated with a given
$(M,g,F)$. Section 3 deals with uniform magnetic fields on
Riemannian surfaces, while the particular case when $(M,g)$ is a
revolution surface is studied in Section 4. In Section 5, a
one-parameter family $\mathcal{F}_m$ of functionals is considered
on an appropriate space of curves $\Lambda$ in the surface. The
Euler-Lagrange equation associated to the variational problem is
then obtained. In Section 6 we define a Gauss-Landau-Hall magnetic
field  on a surface, first in $\R^3$, and then in general. In this
section, we obtain the main result, Theorem \ref{importante},
which asserts that the normal flowlines of a GMF coincide with the
critical points of the appropriate functional $\mathcal{F}_m$.
Stability of the field equation solutions is also studied. In
Section 7, we show a characterization theorem for those revolution
surfaces whose parallels are all normal magnetic curves associated
to a GMF. We close the section studying some particular examples.

\section{Completeness of magnetic curves and more}

An early property of the magnetic curves is the following
conservation's law. Particles evolve with constant speed, and so
constant energy, along the magnetic trajectories

\begin{equation} \label{velocidadconstante}
\frac{d}{dt}\,g(\gamma^{\prime},\gamma^{\prime})=2g(\Phi(\gamma^{\prime}),\gamma^{\prime})=0.
\end{equation}

\vspace{2mm}

In particular, a magnetic curve $\gamma$ is said to be {\it
normal} if it has unit energy, i.e., $\|\gamma'\| \, ^{2} \equiv
1$.

\vspace{2mm}

The existence and uniqueness of geodesics, remains true when one
considers magnetic curves. Thus, for each $p\in M$ and $v\in T_pM$
there is exactly one inextendible (i.e., maximal) magnetic curve,
$\gamma :(-a,a)\longrightarrow M$, of $(M,g,F)$ with $\gamma(0)=p$
and $\gamma^{\prime}(0)=v$, (see for instance \cite{Sachs-Wu}, p.
91). Since the proof of this result does not make use neither the
definiteness of $g$ nor the skew-symmetry of $\Phi$, one has a
{\it present determines the future} type result for an indefinite
metric, Lorentzian in particular, and for any smooth operator.
Even more, the result also works for solutions of a differential
equation that extends that of Landau-Hall in the following terms
\cite{Romero-Sanchez},

\[ \nabla_{\gamma^{\prime}}\gamma^{\prime}=\Phi(\gamma^{\prime})+X
\circ \gamma, \]

\noindent where $X$ is a vector field on a semi-Riemannian
manifold. This setting includes the important case in Mechanics
where $X=-\nabla V$, and $V$ standing for smooth function on $M$,
(\cite{Marsden}, Proposition 3.7.4).

\vspace{2mm}

Nevertheless, the well known homogeneity result for geodesics,
works quite different in non trivial magnetic fields. Therefore,
if $\gamma$ is the inextendible magnetic curve of $(M,g,F)$
determined from the initial data $(p,v)$, the curve $\beta$,
defined by $\beta(t)=\gamma(\lambda t),$ $\lambda
\in\mathbb{R}\backslash\{0\}$, is a magnetic trajectory of
$(M,g,\lambda F)$ and also, when $\lambda>0$, of $(M,(1/\lambda)\,
g,F)$, in both cases determined from initial data $(p,\lambda v)$.
Furthermore, the whole families of magnetic curves of $(M,g,F)$
and $(M,\lambda g,\lambda F)$ coincides, for any constant
$\lambda>0$. Consequently, we have

\begin{prop} \label{homogeneizacion} Let $F$ be a non trivial
magnetic field on a Riemannian manifold, $(M,g)$. Then, there
exists no affine connection on $M$ whose geodesics are the
magnetic curves of $(M,g,F)$.
\end{prop}

\vspace{2mm}

A magnetic field $(M,g,F)$ with Lorentz force $\Phi$, provides, in
a similar way as in \cite{O'Neill}, Prop. 3.28, with a unique
vector field $Q_{\Phi}$ on the tangent bundle $TM$. This is
defined to have integral curves being the lifting to $TM$ of the
magnetic curves, that is, $t \mapsto (\gamma(t),
\gamma^{\prime}(t))$, where $\gamma$ is a magnetic curve of
$(M,g,F)$ (compare with \cite{Lopez}). Certainly this vector field
is nothing but the geodesic flow when $F=0$. Once more, neither
the definiteness of $g$ nor the skew-symmetry of $\Phi$ is needed
to define $Q_{\Phi},$ \cite{Romero-Sanchez}. On the other hand,
the fact that any integral curve of $Q_{\Phi}$ is the velocity of
its projection on $M$, allows us  to think of $Q_{\Phi}$ as a nice
example of the classically so-called {\it second order
differential equation} on $M$. Because the comment previous to
Proposition \ref{homogeneizacion}, $Q_{\Phi}$ is not an {\it
spray}, in general.

A dynamical system with complete trajectories is often thought in
Physics to be persisting eternally. But in many circumstances one
has to deal with incompleteness. So, because of its importance, we
next give criteria to assert when it holds true. An important tool
to study the completeness of the inextendible magnetic curves,
i.e., under what assumptions all the inextendible magnetic curves
are defined on all $\mathbb{R}$, is the vector field $Q_{\Phi}$.
By using Lemma 1.56 in \cite{O'Neill}, it is easily seen the
following result.

\vspace{4mm}

\begin{prop}\label{extendibilidad}
Let $(M,g)$ be a Riemannian manifold, $F$ a magnetic field on $M$
and $\gamma : [a,b) \longrightarrow M$, $a<b$, a magnetic curve of
$F$. The following are equivalent:

\vspace{2mm}

(a) $\gamma$ is extendible to $b$ as a magnetic curve.

\vspace{3mm}

(b) There exists a sequence $\{t_n\} \longrightarrow b$, $t_n \in
[a,b)$ such that the sequence of velocities $\{\gamma '(t_n)\} $
converges in $TM$.
\end{prop}

\vspace{4mm}

Accordingly, a magnetic curve $\gamma : (a,b) \longrightarrow M,
\, a,b\in \mathbb{R}, \, a<b$, of $(M,g,F)$ can be extended to
some open interval $I, \, (a,b)\subset I,$ if and only if $\gamma
(a,b)$ is contained in a compact subset of $M$. Therefore, we get
(compare with Theorem 2.1.18 in \cite{Marsden})

\vspace{4mm}

\begin{prop}\label{nueva}
Let $\gamma$ be an inextendible magnetic curve of $(M,g,F)$ such
that $\gamma(a,b)$ lies in a compact subset of $M$, for every
finite interval $(a,b)$ in its domain. Then, $\gamma$ must be
complete.
\end{prop}

\vspace{2mm}

In particular, if $M$ is assumed to be compact, then we get that
any inextendible magnetic curve of $(M,g,F)$ must be complete.
This fact can be also obtained as a consequence of Corollary
\ref{completitud}, and it will be stated in Remark
\ref{remarkprimera} (a), from a different approach.

\vspace{2mm}

Now, let $\gamma : [a,b) \longrightarrow M$ be a magnetic curve.
Its length $L(\gamma)$ satisfies $L(\gamma) \leq (b-a)\sqrt{e}$,
where $e$ is the (constant) energy of $\gamma.$ For each $t \in
[a,b)$, the distance between $\gamma(a)$ and $\gamma(t)$ satisfies
$d(\gamma(a),\gamma(t)) \leq L(\gamma\mid_{[a,t]}) \leq
(b-a)\sqrt{e}$, which shows that $\gamma([a,b))$ is contained in
the closed metric ball $B$ centered at $\gamma(a)$ and with radius
$(b-a)\sqrt{e}$. Therefore,
$$\gamma^{\prime}([a,b))\subset \{(p,v)\in TM \, : \, p\in B, \,
g(v,v)=e \}\subset TM.$$

Then, we have,

\begin{cor}\label{completitud}
Let $F$ be any magnetic field on a geodesically complete
Riemannian manifold $(M,g)$. Then, all the inextendible magnetic
curves of $(M,g,F)$ are complete.
\end{cor}

\noindent {\bf Proof}. If $(M,g)$ is assumed to be geodesically
complete, then the Hopf-Rinow theorem implies that $B$ must be
compact. Hence $\{(p,v)\in TM \, : \, p\in B, \, g(v,v)=e \}$ is a
compact subset of $TM$. Take now a sequence $\{t_n\}
\longrightarrow b$, $t_n \in [a,b)$, then $ \{ \gamma '(t_n)\}$
lies in a compact subset of $TM$. So, by passing to a subsequence
of $\{t_n\}$, we are under the assumption (b) of Proposition
\ref{extendibilidad}, concluding that $\gamma$ is extendible to
$b$ as a magnetic curve.

\vspace{2mm}

\begin{rem} \label{remarkprimera} {\rm

\begin{description}

\item{}

\item{(a):}  If $M$ is assumed to be compact (therefore $(M,g)$ is
geodesically complete for any Riemannian metric $g$ on $M$), then
we can give an alternative proof of Corollary \ref{completitud}.
In fact, the previous conservation's law [Eq.
(\ref{velocidadconstante})] for the length of velocity vectors of
magnetic curves, implies that the vector field $Q_{\Phi}$ on $TM$
can be restricted to each spherical tangent bundle $U_eM=
\{(p,v)\in TM \, : \,  g(v,v)=e \}\subset TM, \, e>0$. But $U_eM$
is compact whenever $M$ is compact, and hence the restriction of
$Q_{\Phi}$ to $U_eM$ is a complete vector field. This proves that
all the inextendible magnetic curves of $F$ are complete.

\item{(b):}  Proposition \ref{homogeneizacion} has shown a
remarkable difference between magnetic curves and geodesics. The
following non-connecteness fact complement that result. Let us
consider the unit $2$-sphere $\mathbb{S}^{2}(1)$ endowed with its
standard round metric $g,$ and let $F$ be the magnetic field
$F=\mu\,\Omega _2,$ where $\Omega_2$ is the area $2$-form and $\mu
\in \mathbb{R}, \; \mu \neq 0.$ As we will show later (see the
comment after Proposition \ref{proposicionconstante}), the
associated magnetic curves with energy  $e$ are  circles on
$\mathbb{S}^2(1)$ with radius $r=[1+(\mu^2 / e)]^{-1/2}.$ Then, as
$r <1,$ any two antipodal points can not be connected by a
magnetic curve of $(\mathbb{S}^2,g,F).$ Moreover, for any $p\in
\mathbb{S} ^2$, all the inextendible magnetic curves $\gamma$ of
$(\mathbb{S} ^2 \setminus \{-p\},g,F)$ such that $\gamma (0)=p$
are complete.

\item{(c):}  Let  $(M,g)$ be a Riemannian manifold where $g$ is an
incomplete metric. If $F$ is a magnetic field on $(M,g),$ then
there exists a pointwise conformal metric $f^2g$ such that the
inextendible magnetic curves of $(M,f^2g,F)$ are complete. In
fact, there exists $f\in C^{\infty}(M), \; f>0,$ such that $f^2g$
is geodesically complete, \cite{Nomizu-Ozeki}. Therefore,
Proposition \ref{completitud} gives that the magnetic curves of
$(M,f^2g,F)$ are complete.

\item{(d):} It should be observed that the closedness assumption
on the $2$-form $F$ in Proposition \ref{completitud} was not used.
On the other hand, the skew-symmetry of the tensor field $F$ has
played a crucial role [recall the conservation's law
(\ref{velocidadconstante})]. In fact, consider the tensor field
$F=-2x\,dx^2$ on the Euclidean plane \linebreak
$(\mathbb{R}^2,g^0=dx^2+dy^2)$. If $\Phi$ denotes the operator
defined from $F$ using Eq. (\ref{operator}), then $\Phi(\partial /
\partial x)=-2x\, (\partial /\partial x)$ and
$\Phi(\partial / \partial y)= 0$. Therefore,
$\gamma(t)=(x(t),y(t))$ satisfies Eq. (LH) if and only if
$x''(t)+2\,x(t)x'(t)=0$ and $y''(t)=0$. So, $\gamma(t)=(1/t,t)$ is
an inextendible incomplete trajectory of $(\mathbb{R}^2,g^0,F).$

\item{(e):} Finally, let us point out that Proposition \ref{completitud}
cannot be also extended to the indefinite case. In fact, consider
$\R^2$ endowed with the Lorentzian metric \linebreak
$g_L=dx^2-dy^2,$ and define the magnetic field $F=-xdx\wedge dy.$
A curve $(x(t),y(t))$ is a magnetic curve of $(\R ^2,g_L,F)$ if
and only if it satisfies \linebreak $x''(t)=x(t)y'(t), \;\,
y''(t)=x(t)x'(t).$ Then, $\gamma (t)=(2/t,-2/t)$ is an
inextendible magnetic curve which is defined on $(0,\infty).$

\end{description}
}
\end{rem}

\section{Uniform magnetic fields}

From now on, $M$ will be an oriented Riemannian surface with
standard complex structure $J,$ and area element $\Omega_{2}$ so
that $ \Omega_{2}(X,JX)=1$ for any unit vector field $X$ in $M$.

\vspace{2mm}

Given a curve $\gamma$ in $M$ such that
$g(\gamma^{\prime},\gamma^{\prime})=e>0$ is constant, its Frenet
apparatus is $\{T=(1 /\sqrt{e})\,\gamma^{\prime},N=JT\}$. If
$\kappa$ denotes the curvature function, we have the following
well-known Frenet equations

\[ \nabla_{\gamma^{\prime}} T  =  \kappa \sqrt{e}\,N, \qquad
\nabla_{\gamma^{\prime}} N  = -\kappa \sqrt{e}\,T.\]

\vspace{2mm}

Obviously, any magnetic field on a surface, $M$, is determined
from a smooth function, $f$ (the {\it strength}), by $F=f \,
\Omega_{2}$. Therefore, the matrix of $\Phi$ in any orthonormal
frame, $\{X,JX\}$ is given by

\[\left(
\begin{array}{rr}
  0 & -f \\
  f & 0
\end{array}
\right).
\]
In particular, along a magnetic curve $\gamma$ of $(M,g,F)$, with
energy $e$, and relative to its Frenet frame, the Lorentz force is
obtained to be

\[\left(
\begin{array}{rrr}
  0 & -\kappa \, \sqrt{e}\\
  \kappa \,\sqrt{e} & 0
\end{array}
\right).
\]

\noindent Therefore, we get,

\begin{prop}\label{magneticasdeenergiae} The curvature of the  magnetic
curves with energy $e$ is given by $\kappa=f/\sqrt{e}.$ So, the
curvature of the normal magnetics curves completely determines the
Lorentz force, i.e., $f=\kappa$ along these flowlines.
\end{prop}

A parallel magnetic field $F$, i.e., a magnetic field with
constant strength $f=\mu$, is called a {\it uniform magnetic
field}. This class of magnetic fields has been extensively
considered in the literature from different points of view
(\cite{Lopez},\cite{Adachi}-\cite{Wojtkowski}, etc.). The
geo\-metric partner of the Landau-Hall problem, for uniform
magnetic fields, is nothing but the computation of curves with
constant curvature. To be precise, we have,

\begin{prop} \label{proposicionconstante} Let $F=\mu \, \Omega_{2}$ be
a uniform magnetic field, with constant strength $\mu$, on a
Riemannian surface $(M,g)$. A curve $\gamma$ in $M$, with constant
energy $e$, is a magnetic curve of $(M,g,F)$ if and only if it has
constant curvature $\kappa=\mu /\sqrt{e}$.
\end{prop}

On surfaces of constant Gauss curvature, the feature of the normal
flowlines of a non-trivial uniform magnetic field $F=\mu
\,\Omega_2$ is well-known for any uniform magnetic field. On the
Euclidean plane, $\mathbb{R}^{2}$, they are circles with radius
$1/|\mu|$. On the $2$-sphere of radius $r$, $\mathbb{S}^{2}(r)$,
flowlines with energy $e$ are circles with radius
$(r\sqrt{e})/\sqrt{e+r^2\mu ^2}$ ($<r$). In these two backgrounds,
the flowlines are always closed.

On the other hand, the situation in a hyperbolic plane is quite
different. Let $\mathbb{H}^{2}(-G)$ be the upper half-plane (in
$\mathbb{R}^{2}$) endowed with the Lobatchevski metric of
curvature $-G$, $G>0$, that is, the Poincar\'{e} plane. We use
Proposition \ref{proposicionconstante} joint the basic knowledge
of the curves of constant curvature in $\mathbb{H}^{2}(-G)$ (see
any basic text of Riemannian geometry) to make trivial the
following description of the flowlines which is due to A. Comtet,
\cite{Comtet}, and has been mentioned along a large list of
references. The behaviour of normal magnetic curves changes
according to the ratio between the strength, $\mu$, and the
curvature of $\mathbb{H}^{2}(-G)$. Namely,
\begin{itemize}
    \item If $|\mu| / \sqrt{G}>1$, then the trajectories are geodesic circles,
    and therefore they are closed curves.
    \item If $|\mu| / \sqrt{G} \leq 1$, then the trajectories are non-closed curves which
    intersect the boundary line, $\partial\mathbb{H}^{2}(-G)$, of the upper half-plane. In
    particular, they are tangent to this boundary, and so they are horocycles  when
    $|\mu|=\sqrt{G}$.
\end{itemize}


\begin{rem} {\rm
\begin{description}
\item{}

\item{(a):} Let $\gamma$ be  a  curve  with constant geodesic
curvature $\kappa \neq 0$ in any of the three previous constant
curvature surfaces. Then, for a given uniform magnetic field
$F=\mu \,\Omega_2,$ a suitable fitting of the constant speed (and
hence, the energy) of $\gamma$ makes this curve to be a magnetic
curve of $F.$

\item{(b):} Let $(M,g)$ be again one of the three above space forms
and $F=\mu \,\Omega _2$ a uniform magnetic field on $(M,g).$ Then,
any magnetic curve $\gamma$ with energy $e$ of $(M,g,F)$ can be
then considered as a {\it normal} magnetic curve of $\big(
M,(1/e)\, g,(1/e)\, F\big)$ (see the comment previous to
Proposition \ref{homogeneizacion}).
 \end{description}
 }
\end{rem}

\section{The Landau-Hall problem in a surface of revolution}

Let $\alpha(s)=\left(f(s),h(s)\right)$,\, $a<s<b$,\, $f(s)>0$, be
a parametrization by the arclength of a curve, $\mathrm{C}$,
contained in the $\{xz\}$-plane of $\R ^3$. We rotate $\mathrm{C}$
around the $z$-axis to obtain a surface of revolution, say
$M_{\alpha}$, with canonical parametrization in $\mathbb{R}^{3}$
\begin{equation}\label{revolucion}
X(s,v)=\left(f(s)\cos v, f(s)\sin v, h(s)\right), \qquad 0\leq
v\leq 2\pi.
\end{equation} Of course we consider that $M_{\alpha}$
is endowed with the induced metric $g$ of the Euclidean one of
$\mathbb{R}^{3}$.

\vspace{2mm}

Each point of $\mathrm{C}$ describes a parallel, $\gamma_{s}$,
which can be parametrized by arclength in the following way

\[ \gamma_{s}(t)=\left(\, f(s)\cos{\frac{t}{f(s)}}, \; f(s)\sin{\frac{t}{f(s)}}, \; h(s)\,\right), \]
where $0\leq t\leq 2\pi f(s)$.

\vspace{2mm}

The curvature, $\kappa_{s}$, of $\gamma_{s}$ in $M_{\alpha}$, is
computed to be

\[ \kappa_{s}(t)=\|\nabla_{T_{s}} T_{s}\|=\frac{f^{\prime}(s)}{f(s)}, \]
where $T_{s}=\gamma^{\prime}_{s}$ and $\nabla$ is the Levi-Civita
connection of $M_{\alpha}$. In particular, $\kappa_{s}$ is
constant along $\gamma_{s},$ and so this curve is a good candidate
to be a flowline of a suitable uniform magnetic field on
$M_{\alpha}$.

\vspace{2mm}

Let $F=\mu \, \Omega_{2}$ be a uniform magnetic field on
$M_{\alpha}$ with constant strength $\mu$. Then $\gamma_{s}$ is a
normal magnetic flowline of $(M_{\alpha},g,F=\mu \, \Omega_{2})$
if and only if $\kappa_{s}=\mu$ (Proposition
\ref{proposicionconstante}). Therefore, the set of magnetic
parallels of $(M_{\alpha},g,F=\mu \, \Omega_{2})$ can be
identified with the following subset of the interval $(a,b)$

\[  \Gamma_{\mu}=\{s\in(a,b) \, : \, f^{\prime}(s)=\mu f(s)\}. \]

To determine those surfaces of revolution whose parallels are all
normal magnetic curves of a given uniform magnetic field (that is,
those with $\Gamma_{\mu}=(a,b)$) we need to solve the ordinary
differential equation

\[ f^{\prime}(s)=\mu f(s). \]
Obviously, we have two possibilities. The trivial one,
corresponding with the case of a trivial magnetic field (the
strength vanishes), the flowlines are then geodesics, and the
surface of revolution is a right circular cylinder. Otherwise,
since the Gauss curvature of a surface of revolution (in the
canonical parametrization) is given by
\begin{equation}\label{Gaussrevolution}
G(s,t)=-\frac{f^{\prime\prime}(s)}{f(s)},
\end{equation}
we get that $G(s,t)=-\mu ^{2}$, an hence the surface has constant
negative curvature. In particular, we have,

\begin{prop}\label{uniformes}
The parallels of a surface of revolution, $M_{\alpha}$, are all
normal magnetic flowlines of a uniform magnetic field, $F=\mu \,
\Omega_{2}$, if and only if either:
\begin{enumerate}
    \item $M_{\alpha}$ is a right circular cylinder (when $\mu=0$), or
    \item $M_{\alpha}$ is a bugle surface with Gaussian curvature $-\mu ^{2}$.
\end{enumerate}
\end{prop}

\vspace{2mm}

Let us consider the torus of revolution, $\mathbf{T}(r,R)$,
obtained by rotating the circle, $\mathbf{C}$, centered at
$(R,0,0)$ and with radius $r$, ($R>r$), around the $z$-axis. The
circle can be arclength parametrized by $\alpha(s)=\left(R+r \cos
(s/r), \, 0, \, r \sin (s/r)\right)$. Therefore, $f(s)=R+r \cos
(s/r)$, with $0\leq s\leq 2\pi r$.

\vspace{2mm}

Given a uniform magnetic field $F=\mu \, \Omega_{2}$ on
$\mathbf{T}(r,R)$, the set of normal magnetic parallels is
identified to

\[ \Gamma_{\mu}=\{s\in [0,2\pi r] \, : \, H_{\mu}(s)=0\},\]

\noindent where

\[ H_{\mu}(s)=R\mu\,+\,r\mu\cos \,(\frac{s}{r}) \,+\,\sin \,(\frac{s}{r}). \]

\noindent To study $\Gamma_{\mu}=H_{\mu}^{-1}(0)$, we use
elemental calculus. First, we assume that $\mu \neq 0$, otherwise
$\Gamma_{0}=H_{0}^{-1}(0)$ is made up of the two parallels that
are geodesics in $\mathbf{T}(r,R)$. In this setting, we see that
$H_{\mu}$ has exactly two critical points on $\mathbf{C}$ (the
maximum and the minimum) which are antipodal. In fact,
$H_{\mu}^{\prime}(s)=0$ if and only if $\cot (s/r)=r\mu$ and it
happens just in the following two antipodal points

\begin{eqnarray*}
    p_{\mu} & = & \left(\,\cos{\frac{s}{r}}, \; \sin{\frac{s}{r}}\,\right)=\left(\frac{r \mu}{\sqrt{1+r^{2}\mu^{2}}}, \, \frac{1}{\sqrt{1+r^{2}\mu^{2}}}\right),\\
    q_{\mu} & = & \left(\,\cos{\frac{s}{r}}, \; \sin{\frac{s}{r}}\,\right)=-\left(\frac{r \mu}{\sqrt{1+r^{2}\mu^{2}}}, \,
    \frac{1}{\sqrt{1+r^{2}\mu^{2}}}\right),
\end{eqnarray*}

\noindent The values of $H_{\mu}$ in $p_{\mu}$ and $q_{\mu}$ are

\begin{eqnarray*}
H_{\mu}(p_{\mu}) & = & R \mu+\sqrt{1+r^{2}\mu^{2}}; \qquad {\rm the \, maximum \, of} \quad H_{\mu}, \\
H_{\mu}(q_{\mu}) & = & R \mu-\sqrt{1+r^{2}\mu^{2}}; \qquad {\rm
the \, minimum \, of} \quad H_{\mu}.
\end{eqnarray*}

\noindent We call $D_{\mu}$ the diameter in $\mathbf{C}$
determined by $p_{\mu}$ and $q_{\mu},$ and let
$\rho=(R^{2}-r^{2})^{-1/2}.$ We distinguish two cases.

\medskip

\noindent (A) If $\mu>0$, then the point $p_{\mu}$, where
$H_{\mu}$ gets its maximum, lies in the first quadrant of the
circle $\mathbf{C}$. Since $H_{\mu}(p_{\mu})>0$, then
$\Gamma_{\mu}\neq\emptyset$ if and only if $H_{\mu}(q_{\mu})\leq
0$ (the minimum is non positive) that is $\mu \leq \rho$.
Certainly if the equality holds, then $H_{\rho}$ vanishes only at
$q_{\rho}=\left(-r/R, \, -1/(\rho R)\right)$. Otherwise, $H_{\mu}$
vanishes exactly in two points, say $z_{\mu}$ and $w_{\mu}$, which
are separated by $D_{\mu}$ so they lie in different half circles.

\medskip

\noindent (B) If $\mu<0$, then the point $p_{\mu}$, where
$H_{\mu}$ gets its maximum, lies in the second quadrant of the
circle $\mathbf{C}$. Since the minimum is negative,
$H_{\mu}(q_{\mu})<0$, then $\Gamma_{\mu}\neq\emptyset$ if and only
if $H_{\mu}(p_{\mu})\geq 0$ and it happens if and only if $\mu
\geq -\rho$. It is clear that when the equality holds, then
$H_{\rho}$ vanishes just at $p_{\rho}=\left(-r/R, \, 1/(\rho
R)\right)$. However, if $\mu>-\rho$, then $H_{\mu}$ has exactly
two zeroes, say $z_{\mu}^{\prime}$ and $w_{\mu}^{\prime}$, which
are obviously separated by $D_{\mu}$ so they lie in different half
circles.

\medskip

All this information can be summarized in the following statement,

\begin{prop} \label{proptoro} Let $F=\mu \,\Omega_{2}$ be a uniform
magnetic field on a torus of revolution, $\mathbf{T}(r,R)$. Then
$\left(\mathbf{T}(r,R), \,g, \mu \, \Omega_{2}\right)$ has normal
magnetic parallels if and only if \break $\mu \in \left[-\rho, \,
\rho \right]$. Furthermore,
\begin{enumerate}
\item If $\mu=-\rho,$ then there is one
normal magnetic parallel obtained by rotating the point $\left(
R+(r^2/R),r/(\rho R) \right)\in \mathbf{C}$.
\item If $\mu=\rho$, then there is one
normal magnetic parallel obtained by rotating the point $\left(
R-(r^2/R),-r/(\rho R) \right)\in \mathbf{C}$.
\item If $\mu \in
\left(-\rho , \, \rho \right)$, then $H_{\mu}$ has exactly two
normal magnetic parallels obtained by rotating two points of
$\mathbf{C}$ that are separated by $D_{\mu}$.
\end{enumerate}
\end{prop}

\begin{ex} \label{catenoide} {\rm  Let $C$ be the cathenoid generated by
revolving the cathenary curve $\alpha (t)=(\cosh t,0,t), \; t\in
\mathbb{R}$ around the  $z$-axis. Let $F$ be a uniform magnetic
field with constant strength $\mu \neq 0$, defined on $C$. Then,
it can be shown that there exist magnetic parallels if and only if
$\mu \in \left[ -1/2,1/2 \right]$. Moreover, we have:

\begin{enumerate}

\item If $\mu=1/2$ (resp. $\mu=-1/2$), there exists
only a normal magnetic parallel corresponding to $t=\ln
(1+\sqrt{2})$ (resp. $t=-\ln (1+\sqrt{2})$).
\item If $\mu \in \left( -1/2,0\right)$ (resp. $\mu \in \left(
0,1/2\right))$, then $C$ has two normal magnetic parallels which
are located in the region $t>0$ (resp. $t<0$).

\end{enumerate}}
\end{ex}

\begin{ex} \label{cicloide} {\rm  Let $S$ be the revolution surface obtained by rotating the
cycloid curve $\alpha (t)=(a(1-\cos t),0,a(t-\sin t)), \; t\in
(0,2\pi)$ around the $z$-axis. It is easy to see:

\begin{enumerate}
\item If $\mu >0$, then $S$ has only a normal magnetic parallel
corresponding to the parameter value $t_0=2\arccos \, \left[
\left( -1+\sqrt{1+16a^2\mu^2}\right) /(4a\mu)\right].$
\item If $\mu<0$, then $S$ has also only a normal magnetic parallel
corresponding to $t_1=2\pi -t_0$.
\end{enumerate}}
\end{ex}

\begin{ex} {\rm  Finally, let us consider the cone $M$ generated by the line
$\alpha (t)=(at,0,bt), \; a^2+b^2=1, \; t>0$ around the $z$-axis.
Then, it can be seen that for any $\mu>0$, there exists a unique
normal magnetic parallel given by $t=1/ \mu$.}
\end{ex}

\section{Relativistic particles with rigidity of order one}

The search for Lagrangians describing spinning particles (both
massive and massless) has a long history. An interesting and
unconventional approach is to provide the necessary extra degrees
of freedom by actions whose densities depend on higher order
geometrical invariants. In particular, this means that those extra
bosonic variables must be encoded in the geometry of the world
trajectories. The simplest models are those involving density
Lagrangians that depend on the curvature, $\kappa$, of the
worldlines (\cite{Arroyo}-\cite{Nesterenko1}, etc.). In
particular, actions that depend linearly from $\kappa$
(\cite{Plyushchay2},\cite{Arroyo-Barros-Garay}-\cite{Plyushchay3},
etc.) will be considered in this section. These models describing
a massive relativistic boson \cite{Plyushchay}.

\vspace{2mm}

Suppose that $\Lambda$ is a suitable space of curves (closed
curves or clamped curves, for instance) in a Riemannian surface
$(M,g)$. Define a one-parameter family of functionals
$\mathcal{F}_{m}:\Lambda\rightarrow\mathbb{R}, \; m\in \R,$ by
\begin{equation}\label{funcional}
\mathcal{F}_{m}(\gamma)=\int_{\gamma}(\kappa+m)ds,
\end{equation}where $s$ stands for the arclength parameter of curves
$\gamma\in\Lambda$. In order to obtain the first variation of
these actions, we use the following standard machinery (see for
instance \cite{Barros}). For a curve $\gamma:[0,L]\rightarrow M$,
we take variations
$\Theta=\Theta(t,r):[0,L]\times(-\varepsilon,\varepsilon)\rightarrow
M$ with $\Theta(t,0)=\gamma(t)$. Then, we have the variation
vector field $W=W(t)=\left( \partial\Theta /\partial
r\right)(t,0)$ along the curve $\gamma$. We also put
$V=V(t,r)=\left( \partial\Theta /\partial r\right)(t,r)$,
$W=W(t,r)$, $v=v(t,r)=\|V(t,r)\|$, $T=T(t,r)$, $N=N(t,r)$, with
the obvious meanings. The corresponding reparametrizations will be
denoted by $V(s,r)$, $W(s,r)$ etc. The variations of $v$ and
$\kappa$ in $\gamma$, in the direction of $W$, can be obtained to
be

\begin{equation}\label{v}
W(v)  =  g(\nabla_{T} W,T)\,v,
\end{equation}
\begin{equation}\label{curvature}
W(\kappa)  =  g(\nabla_{T}^{2} W,N)-2g(\nabla_{T} W,T)\kappa+G \,
g(W,N),
\end{equation}
here $G$ denotes the Gauss curvature of $(M,g)$ and $\nabla$ its
Levi-Civita connection.

\vspace{2mm}

Now, we use a standard argument that involves the above formulas
so as some integrations by parts to obtain the first derivative of
$\mathcal{F}_{m}$
\begin{equation}\label{variacion}
\delta\mathcal{F}_{m}(\gamma)[W]=\int_{\gamma}
g\left(\Omega(\gamma),W\right)
ds+\left[\mathcal{B}(\gamma,W)\right]_{0}^{L},
\end{equation}where $\Omega(\gamma)$ and $\mathcal{B}(\gamma,W)$ denote the
Euler-Lagrange and the boundary operators and they are
respectively given by

\begin{eqnarray*}
\Omega(\gamma) & = & (G-m \, \kappa)N, \\
\mathcal{B}(\gamma,W) & = & g(\nabla_{T} W,N)+m \, g(W,T).
\end{eqnarray*}

\begin{prop}\label{clamped} {\rm (Clamped curves)} Given points $q_{1}, \, q_{2}\in M$ and unit vectors $x_{1}\in
T_{q_{1}}M$ and $x_{2}\in T_{q_{2}}M$, define the space of curves
\[ \Lambda=\{\gamma:[t_{1},t_{2}]\rightarrow M \, : \,
\gamma(t_{i})=q_{i}, \, T(t_{i})=x_{i}, \, N(t_{i})=Jx_{i}, \,
1\leq i\leq 2\}. \] Then, the critical points of the functional
$\mathcal{F}_{m}:\Lambda\rightarrow\mathbb{R}$ are characterized
by the following Euler-Lagrange equation

\[   G\mid_{\gamma}=m \, \kappa,    \]
where $G\mid _{\gamma}$ denotes the Gauss curvature of $(M,g)$
along $\gamma$.
\end{prop}

\noindent {\bf Proof.} Let $\gamma\in\Lambda$ and $W\in
T_{\gamma}\Lambda$, then $W$ defines a curve in $\Lambda$
associated with a variation $\Theta$ of $\gamma$. We can perform
the following computations along $\Theta$

\begin{eqnarray*}
W & = & d\Theta(\partial_{r}), \\
\nabla_{T} W & = & f \, T+d\Theta(\partial_{r} T),
\end{eqnarray*}
where $f=\partial_{r}(\ln \, v)$. We evaluate these formulas along
$\gamma$ by making $r=0$ and use that $\Theta$ is a curve in the
space $\Lambda$ to obtain

\begin{eqnarray*}
W(t_{i}) & = & 0, \\
\nabla_{T} W(t_{i}) & = & f(t_{i})x_{i}.
\end{eqnarray*}
As a consequence,
$\left[\mathcal{B}(\gamma,W)\right]_{t_{1}}^{t_{2}}=0$. So, using
Eq. (\ref{variacion}), we have that $\gamma$ is a critical point
of $\mathcal{F}_{m}:\Lambda\rightarrow\mathbb{R}$, that is
$\delta\mathcal{F}_{m}(\gamma)[W]=0$ for any $W\in
T_{\gamma}\Lambda$, if and only if $\Omega(\gamma)=0$ which proves
the statement.


\vspace{4mm}

Similarly, we can obtain the following

\begin{prop}\label{closed} {\rm (Closed curves)} Let $\mathcal{C}$ be the space of immersed closed curves in
$(M,g)$. The critical points of the functional
$\mathcal{F}_{m}:\mathcal{C}\rightarrow\mathbb{R}$ are those
closed curves that are solutions of the following Euler-Lagrange
equation

\[  G\mid _{\gamma}=m \, \kappa.   \]
\end{prop}

\section{Gaussian magnetic fields}

Let $M$ be a surface immersed in the Euclidean three-space,
$\mathbb{R}^{3}$, so the metric, $g$ is the induced one. We denote
by $N : M \longrightarrow \mathbb{S}^{2}$ its Gauss map and
$d\sigma^{2}$ will stand for the area element on the unit round
sphere $\mathbb{S}^{2}$. The two form $N^{*}(d\sigma^{2})$ on $M$
can be used, for example, to measure areas of the spherical images
or topological total charges of solitons in the $O(3)$ non-linear
sigma model (see for instance \cite{Ody}, \cite{Tsurumaru} and
references therein). In this section we will consider magnetic
fields of the type

\[  F=\frac{1}{m} N^{*}(d\sigma^{2}), \]
where $m$ is a non zero constant. We call then {\it Gaussian
magnetic fields} (GMF). It is well known that
$N^{*}(d\sigma^{2})=G \, \Omega_{2}$, $G$ denoting the Gaussian
curvature of $(M,g)$ and this, in particular, implies that we can
consider these kind of magnetic fields with no mention to the
surrounding space. Namely, a GMF is always of the type
\begin{equation}\label{Gaussian}
F=\frac{G}{m}\,\Omega_2.
\end{equation}

The Lorentz force of a GMF is computed to be $\Phi=(G/m) \, J$,
where $J$ is the standard complex structure in $M$. In particular,
for any unit vector field, $X$, on $M$, the matrix of $\Phi$, in
the terminology of Section 3, with respect to an orthonormal frame
$\{X,JX\}$ is given by

\[\left(
\begin{array}{rr}
  0 & - \frac{G}{m} \\
   \frac{G}{m} & 0
\end{array}
\right).
\]

\vspace{2mm}

In this framework, we can combine Proposition
\ref{magneticasdeenergiae} and the field equations of the particle
models defined from $\mathcal{F}_{m}$, see Eq. (\ref{funcional}),
to obtain the following amazing relationship between the flow of a
GMF and the worldline trajectories of relativistic particles with
order one. To be precise, we have,

\begin{thm}\label{importante}
Let $\gamma\in \Lambda$ be a curve {\rm (clamped or closed)} in
$(M,g)$. Then it is a normal flowline of $\left( M,g,F=(G/m) \,
\Omega_{2} \right)$ if and only if it is a critical point {\rm
(world line)} of the action
$\mathcal{F}_{m}:\Lambda\rightarrow\mathbb{R}$ given by

\[ \mathcal{F}_{m}(\gamma)=\int_{\gamma} (\kappa+m) \, ds. \]

\end{thm}

\noindent At this point, we can take advantage of the variational
approach to study stability of the GMF flowlines. Therefore, we
need the second derivative of $\mathcal{F}_{m}$ in a critical
point, say $\gamma$ in a suitable space of curves, $\Lambda$
(recall closed or clamped curves). After some long computations
(see \cite{Arroyo} for details) one can obtain the following
expression

\begin{equation}\label{segundavariacion}
\delta^{2}\mathcal{F}_{m}(\gamma)[W]=\int_{\gamma}g\left(W,\nabla_{W}
\Psi\right)|_{\gamma} \, ds,
\end{equation}where $\Psi$ denotes
the vector field, along a variation of $\gamma$, given by
$\Psi=(G-m \, \kappa) \, N$. Now, we choose $W=\Phi \, N$ to
obtain

\[ g\left(W,\nabla_{W} \Psi\right)|_{\gamma}=\Phi^{2} \, N(G-m \, \kappa),   \]
where the right hand term is restricted to $\gamma$. However the
variation of $\kappa$ was given in Eq. (\ref{curvature}); so, in
particular, we have $N(\kappa)=\kappa^{2}+G$. Then, one gets from
Eq. (\ref{segundavariacion})

\[  \delta^{2}\mathcal{F}_{m}(\gamma)[W]=\int_{\gamma}\Phi^{2}\left(N(G)-\frac{1}{m}(G^{2}+m^{2}G)\right) \, ds.  \]

Hence, we have the following useful test of stability.

\begin{prop}\label{caracterizacionestabilidad}
A critical point, $\gamma\in\Lambda$, of $\mathcal{F}_{m}$ is
stable if and only if the function $\; N(G)-\displaystyle (
1/m)(G^{2}+m^{2}G)$ is signed along $\gamma$.
\end{prop}

It should be observed that the previous test has the following
geometrical meaning. Put $\varphi=G-m \, \kappa$, then
$\gamma\subset\varphi^{-1}(0)$, because it is a critical point of
$\mathcal{F}_{m}$, and then stability means that $\gamma$ is made
up of regular points of $\varphi$. Moreover, observe that this
happens if $0$ is a regular value of $\varphi$.

\vspace{2mm}

Now, let us use all this information in the following elemental
setting. We consider $M=\mathbb{S}^{2}(1)$ the unit round sphere.
Then any GMF, $F=(G/m) \, \Omega_{2}=(1/m) \, \Omega_{2}$, is
uniform. However, when studying uniform magnetic fields on a round
sphere, we can not talk about stability of magnetic trajectories.
This is not the case of our approach.

\vspace{2mm}

In this setting, the magnetic curves are, according Theorem
\ref{importante}, the critical points of the functional

\[ \mathcal{F}_{m}(\gamma)=\int_{\gamma}(\kappa+m) \, ds=\int_{\gamma}\kappa \,
ds+m\, L(\gamma),
\]
and they are nothing but those curves that satisfy $1=m \,
\kappa$, that is geodesic circles with geodesic curvature
$\kappa=1/m$.

\vspace{2mm}

On the other hand, we can use the Gauss-Bonnet formula to see that
this variational problem is equivalent to that associated with the
action $\mathcal{D}_{m}:\mathbf{D}\longrightarrow\mathbb{R}$
defined by

\[ \mathcal{D}_{m}(\Delta)=\int_{\Delta} G \, \Omega_{2}+m\int_{\partial\Delta}ds={\rm Area}(\Delta)+m \,
 L(\partial\Delta),
\]
acting on the space $\mathbf(D)$ of simply-connected domains,
$\Delta$ in $\mathbb{S}^{2}$, with the same boundary
$\gamma=\partial\Delta$. This is nothing but the isoperimetric
problem in the round sphere. The solution is a couple of domains
$\Delta_{1}$ and $\Delta_{2}$ (the maximum and the minimum) with
common boundary a geodesic circle, $\gamma$, of curvature
$\kappa=1/m$. Since $N(G)=0$ then, $-(1+m^{2})/m$ has obviously
sign for any choice of the coupling constant $m$. Consequently,
calling to Proposition \ref{caracterizacionestabilidad}, the
solutions are stable.

\vspace{2mm}

The case where $m=0$ deserves a few words. First of all GMF with
$m=0$ could be considered as a limiting case, however, after our
variational approach, it can be identified with the massless
Plyushchay model, \cite{Plyushchay1}, which is governed by the
Lagrangian

\[  \mathcal{F}_{0}(\gamma)=\int_{\gamma}\kappa \, ds.  \]

\noindent This model has been considered with detail in
\cite{Arroyo-Barros-Garay}. For example the sphere does not admit,
non only minima (maxima) for this model but also critical points.
However, an anchor ring has two critical points corresponding to
the two parallels of parabolic points.

\section{GMF flowlines on some non-constant Gauss curvature surfaces}

In this last section we would like to analyze when certain
relevant curves on some non-constant Gauss curvature surfaces are
in fact magnetic.

We recalled the explicit expression of the Gauss curvature,
$G(s,v)$, of a surface of revolution, $M_{\alpha}$, see Eq.
(\ref{Gaussrevolution}). Consequently, we can assert now that a
parallel, $\gamma_{s}$, is a normal flowline of the GMF given by
$(G/m) \, \Omega_{2}$ on $M_{\alpha}$ if and only if
\begin{equation}\label{parallel}
f^{\prime\prime}(s)+m \, f^{\prime}(s)=0.
\end{equation}
\noindent Next, we will obtain the surfaces of revolution whose
parallels are all normal flowline of a GMF. In contrast to the
case of a uniform magnetic field, where only the bugle surface
appeared as a solution (see Proposition \ref{uniformes}), now the
general solution is made up of a three parameter family of
surfaces which includes the bugle surface too.

\begin{thm} The normal flow of a GMF, $(G/m) \,\Omega_{2}$,
in a surface of revolution, $M_{\alpha}$, is invariant under
rotations if and only if the profile curve of $M_{\alpha}$ lies in
the following three parameter family of arclength parametrized
plane curves

\[ \alpha(s)=\left(f(s), \, \int_{0}^{s}\sqrt{1-f^{\prime}(s)^{2}} \, ds\right),         \]
where

\[ f(s)=\frac{1}{m}\big(a+c \, \exp(-m \, s+b)\big); \qquad a,b,c\in \mathbb{R} \quad {\rm with} \quad a>0.   \]
\end{thm}

Observe that the general solution of the ordinary differential
equation (\ref{parallel}) has the form $f(s)=\big(a+c \, \exp(-m
\, s+b)\big)/m$, so the proof of the last result becomes obvious.
Observe also that the above characterized class of surfaces of
revolution includes the bugle surface ($a=0$) and the right
circular cylinder ($c=0$) too.

\vspace{2mm}

It should be noticed the following coupling phenomenon in a
surface of revolution, $M_{\alpha}$, between the GMF, $F_{1}=(G/m)
\, \Omega_{2}$ and the uniform magnetic field $F_{2}=-m \,
\Omega_{2}$, for some values of the coupling constant $m$.
Suppose, for example, that $M_{\alpha}=\mathbf{T}(r,R)$ is a torus
of revolution and $\rho=(R^2-r^2)^{-1/2}$ (notation as in Section
4). Then, $F_{1}$ always has two parallels being normal magnetic
curves, no matter the value of $m$. Now, we use Proposition
\ref{proptoro} to obtain the following statement,

\begin{prop} If \, $-m\in
\left( \rho, \, \rho \right)$, then both $F_{1}$ and $F_{2}$ have
two normal magnetic parallels coming from points alternatively
placed in the profile circle. Moreover they collapse when \, $-m$
goes to $-\rho$ or $\rho$.
\end{prop}

\noindent {\bf Proof.} For any value of $m$ in $\mathbb{R}$,
$(\mathbf{T}(r,R),g,F_{1})$ has two normal magnetic parallels
obtained by rotation of the two antipodal points in $\mathbf{C}$,
defined by $\cot \, (s/r)=-r \, m$. These two points are just
those determining the diameter $D_{-m}$ that separates the two
magnetic parallel of $(\mathbf{T}(r,R),g,F_{2})$ when $-m\in
\left(-\rho, \, \rho \right)$. The second part of this statement
follows similarly when use points 1 and 2 of Proposition
\ref{proptoro}.

\vspace{4mm}

We finish the paper showing several examples.

\vspace{3mm}

\begin{ex} {\rm Let $\beta(s)$ be an arclength
parametrized curve contained in a plane, $\Pi$ (with unit normal
vector $B_{0}$), in $\mathbb{R}^{3}$. We denote by $\{T(s),N(s)\}$
a Frenet frame along $\beta(s)$, so that $T(s)\wedge N(s)=B_{0}$,
and $\kappa(s)$ will stand for its curvature function. For a
suitable $r>0$, we define a tube of radius $r$, say
$\mathrm{T}_{\beta}(r)$, as the surface given by

\[ X(s,v)=\beta(s)+r\big(\cos(v) \, N(s)+\sin(v) \, B_{0}\big). \]

\noindent We denote by $\Lambda_{\beta}=\{\,\gamma_{v}, \, : \,
v\in[0,2\pi] \,\}$ the family of curves in the tube obtained when
we make $v$ constant. The curvature of these curves in
$\mathrm{T}_{\beta}(r)$ can be obtained, from a direct
computation, to be

\[  \kappa_{v}(s)=\frac{\kappa(s) \, \sin(v)}{1-r \, \kappa(s) \, \cos(v)}.    \]
Notice that it is not constant unless $\beta(s)$ is chosen to be
constant curvature.

\vspace{2mm}

On the other hand, the Gauss curvature of the tube
$\mathrm{T}_{\beta}(r)$ is computed to be

\[ G(s,v)=-\frac{\kappa(s) \, \cos(v)}{r\big(1-r\kappa(s) \, \cos(v)\big)}.    \]

\noindent Now, we can apply these formulas together with the
Euler-Lagrange equations associated with the GMF, $F=(G/m) \,
\Omega_{2}$ (Propositions \ref{clamped}, \ref{closed}), to see
that there exist exactly two curves (clamped or closed) in
$\Lambda_{\beta}$ that are normal magnetic trajectories. They are
obtained for $\cot(v)=-r \, m$ and this is, formally, the same
result that we have obtained for a torus of revolution
(Proposition \ref{proptoro}) which can be regarded as a tube
around a circle.}
\end{ex}

\vspace{3mm}

\begin{ex} {\rm Similarly, for a curve, $\beta(s)$, in
$\mathbb{R}^{3}$ with Frenet frame $\{T(s),N(s),B(s)\}$, curvature
$\kappa(s)$ and torsion $\tau(s)$, one can define the tube
$\mathrm{T}_{\beta}(r)$ by

\[ X(s,v)=\beta(s)+r\big(\cos(v) \, N(s)+\sin(v) \, B(s)\big). \]

\noindent In  particular, if $\beta(s)$ is a helix ($\kappa$ and
$\tau$ are both constant) then the curvature function,
$\kappa_{v}(s)$ of the curves in $\Lambda_{\beta}=\{\,\gamma_{v},
\, : \, v\in[0,2\pi]\, \}$ satisfy

\[  \kappa_{v}^{2}=\frac{\kappa^{2} \, \sin^{2}(v)}{\big(1-r \, \kappa \, \cos(v)\big)^{2}+r^{2}\tau^{2}}. \]

\noindent Now, the curves in $\Lambda_{\beta}$ that are normal
flowlines of \, $(G/m) \, \Omega_{2}$ on the helicoidal tube
$\mathrm{T}_{\beta}(r)$ correspond with the zeroes of the function
$\vartheta:\mathbb{S}^{1}\rightarrow\mathbb{R}$ defined by

\[ \vartheta(v)=\big(1-r \, \kappa \, \cos(v)\big)^{2}\big(\cos^{2}(v)-r^{2}m^{2}\sin^{2}(v)\big)+r^{2}\tau^{2}\cos^{2}(v). \]
However, we have

\[  \vartheta(0)=\vartheta(\pi)=(1-r \, \kappa)^{2}+r^{2}\tau^{2}>0 \quad \textrm{and}
\quad  \vartheta(\frac{\pi}{2})=\vartheta(\frac{3\pi}{2})=-r^{2}
m^{2}<0. \] Therefore, there exist four curves of
$\Lambda_{\beta}$ in the flow of \, $(G/m) \, \Omega_{2}$. }
\end{ex}

\vspace{2mm}

\begin{ex} {\rm On the cathenoid (Example \ref{catenoide}), the GMF given by \, $(G/m)\,\Omega_2$ has a
unique normal magnetic parallel for all $m$. If $m>0$, it is
obtained for a $t_0<0$. If $m<0$, then it is obtained for a
$t'_0=-t_0>0$. }
\end{ex}

\vspace{2mm}

\begin{ex}  {\rm On the hyperboloid of revolution obtained from
Eq. (\ref{revolucion}) by putting $f(t)=\cosh t$ and $h(t)=\sinh
t$, the GMF given by \, $(G/m)\,\Omega_2$ has also a unique normal
magnetic parallel for all $m$, analogously to the previous case. }
\end{ex}

\vspace{2mm}

\begin{ex} {\rm  On the cicloidal surface (Example \ref{cicloide}),
the GMF given by \, $(G/m)\,\Omega_2$ has a unique normal magnetic
parallel for $m\in\left(-\infty , -1/(4a)\right)\bigcup\left(
1/(4a),\infty\right)$. If $m>0$, $t_0=\arccos
\left((1/(4am)\right)$, whereas if $m<0$, then $t'_0=2\pi -t_0$.}
\end{ex}

\vspace{3mm}

\noindent{\bf \Large Conclusions}

\vspace{2mm}

Oriented surfaces, $M$, in $\mathbb{R}^{3}$ admit two natural
$2$-forms. First, the area element, $\Omega_{2}$, associated with
the induced metric, $g$. Second, the area element,
$N^{*}(d\sigma^{2})$, of its spherical image under the Gauss map,
$N:M\rightarrow\mathbb{S}^{2}$. It is well known that these
$2$-forms are nicely related by

\vspace{2mm}

\[  N^{*}(d\sigma^{2})=G \, \Omega_{2},    \]

\vspace{2mm}

\noindent where $G$ enotes the Gaussian curvature of $g$. In
particular, both $2$-forms are intrinsic and then they are defined
once we know a Riemannian metric, $g$, on $M$. Associated with
these $2$-forms appear two classes of magnetic fields on $(M,g),$

\begin{enumerate}
    \item The class made up of the constant multiples of the former
    one, $\mathcal{C}_{1}=\{\mu \, \Omega_{2} \, : \, \mu\in\mathbb{R}\}$,
    provides that of uniform magnetic fields, with strength $\mu$,
    on $(M,g)$. The corresponding Landau-Hall problem has been
    widely studied along the literature. Even in this paper, we
    have obtained some new information  relative to uniform
    magnetic field essentially in a surface of revolution. For
    example, we have characterized right circular cylinders and
    bugle surfaces as the only surfaces of revolution whose
    parallels are all normal magnetic flowlines of a uniform
    magnetic fields.
    \item The class of the constant multiples of the later one,
    $\mathcal{C}_{2}=\{\mu \, N^{*}(d\sigma_{2}) \,  : \,
    \mu\in\mathbb{R}\}$, constitutes a class of magnetic fields
    that in this paper are introduced under the terminology of
    Gauss-Landau-Hall magnetic fields (GMF). In this case the
    strength is given by $\mu \, G$ and obviously both classes
    coincide when $(M,g)$ has constant curvature.
\end{enumerate}

\vspace{2mm}

In this paper, we wish to state the importance and nice interest
of GMF on surfaces. In fact, the chief result of the paper appears
when we study the Landau-Hall problem associated with a GMF (which
we call the Gauss-Landau-Hall problem). Then, we are be able to
show that this problem is equivalent to the dynamics of a massive
relativistic boson. This provides an amazing relationship between
two, a priori, quite different physical phenomena.

\vspace{2mm}

Therefore, we can use two points of view to study each of the two
involved problems. On one hand, one can study completeness,
homogeneity and so on, in the dynamical study of bosonic
worldlines. By the way, we have introduced a section with new
results on these topics. But on the other hand, the
Gauss-Landau-Hall problem can be regarded as a variational problem
globally stated. In this setting, flowlines are critical points of
an action which has been used to model relativistic particles with
order one rigidity. In particular, we can talk about, and so we
study, global stability of normal flowlines of a GMF. Say finally
that under this equivalence, the model to describe a massless
particle with arbitrary helicity correspond with a limit case
obtained when the force of the GMF increases arbitrarily.

\vspace{2mm}

We believe that this new point of view in the study of GMF is
physically remarkable and it could be extended to other classes of
magnetic fields.

\bigskip

\vspace{2mm}

\noindent {\bf \Large{Acknowledgments}}

\vspace{2mm}

Research partially supported by MCYT FEDER Grant BFM 2001-2871-C04
and by Acci\'{o}n Coordinada Grupos de Investigaci\'{o}n FQM-324 and
FQM-327, Junta de Andaluc\'ia.

\end{document}